\documentclass{aa} 
\usepackage{caption}
\usepackage{amsmath,xcolor}
\usepackage{lineno}
\usepackage[utf8]{inputenc}
\usepackage[title]{appendix}%
\usepackage{graphicx}
\usepackage[version=3]{mhchem}
\usepackage{txfonts}
\begin{document}

   \title{Multi-transition study of methanol towards NGC 1068 with ALMA}
    \titlerunning{Multi-transition study of methanol towards NGC 1068 with ALMA}
    \authorrunning{Huang et al.}


   \author{K.-Y. Huang
          \inst{1}\fnmsep\thanks{kyhuang@strw.leidenuniv.nl}, 
          D. Abbink
          \inst{1}, 
          S. Viti
          \inst{1}, 
          S. Garc{\'{\i}}a-Burillo
          \inst{2}, 
          }

   \institute{Leiden Observatory, Leiden University, PO Box 9513, 2300 RA Leiden, The Netherlands
         \and
              Observatorio Astron{\'{o}}mico Nacional (OAN-IGN)-Observatorio de Madrid, Alfonso XII, 3, 28014-Madrid, Spain
             }

   \date{Submitted: December 2023; accepted: }

 
  \abstract
   {The outflowing molecular gas in the circumnuclear disk (CND) of the nearby (D=14 Mpc) AGN-starburst composite galaxy NGC 1068 is considered as a manifestation of ongoing AGN feedback. The large spread of velocities from the outflowing gas is likely driving various kinds of shock chemistry across the CND. }
   {We performed a multiline molecular study using \ce{CH3OH} with the aim of characterizing the gas properties probed by \ce{CH3OH} in the CND of NGC 1068, and investigating its potential association with molecular shocks. }
   {Multi-transition \ce{CH3OH} were imaged at the resolution of $0\,.\!\!^{\prime\prime}5-0\,.\!\!^{\prime\prime}8$
   with the Atacama Large Millimeter/submillimeter Array (ALMA). We performed non-LTE radiative transfer analysis coupled with a Bayesian inference process in order to determine the gas properties such as the gas volume density and the gas kinetic temperature. }
   {The gas densities traced by \ce{CH3OH} point to $\sim 10^{6}$ cm\textsuperscript{-3} across all the CND regions. The gas kinetic temperature cannot be well constrained in any of the CND regions though the inferred temperature is likely low ($\lesssim 100$ K).}
   {{The low gas temperature traced by \ce{CH3OH} suggests shocks and subsequent fast cooling as the origin of the observed gas-phase \ce{CH3OH} abundance. We also note that the E-/A- isomer column density ratio inferred is fairly close to unity, which is interestingly different from the Galactic measurements in the literature. It remains inconclusive whether \ce{CH3OH} exclusively traces slow and non-dissociative shocks, or whether the \ce{CH3OH} abundance can actually be boosted in both fast and slow shocks. }}


   \maketitle
%
\section{Introduction}
Feedback from active galactic nuclei (AGN) and star formation activities are both important agents in influencing the interstellar medium (ISM) and the evolution of galaxies. 
Shocks and turbulence can be associated with star formation activities due to protostellar outflows and supernova explosions that affect  both chemistry and kinematics in the ISM. 
These energetics also act as heating sources (mechanical heating) aside from the ionization via X-ray and cosmic rays. 
On the other hand, AGNs are often considered to be X-ray dominated regions (XDRs) and are often thought to have an enhanced cosmic ray ionization rate (CRIR) and radiation field 
\citep[e.g.][]{Meijerink_Spaans_2005,Viti+2014}. 
While converting gravitational potential energy and kinetic energy into radiation, AGN can also trigger jets and the associated multi-phase mass outflows that inject energy into the ISM. 
It is therefore important to build an understanding of the interplay between these two feedback mechanisms. 
In this context, AGN-starburst composite galaxies serve as prime sources to study feedback from both AGN and star formation. 

NGC1068 is a nearby (D = 14 Mpc \citealp{Bland-Hawthorn+1997}, $1'' \sim 70$ pc) Seyfert 2 galaxy and is considered to be the archetype of a composite AGN-starburst system. 
The proximity of this composite galaxy makes it an ideal laboratory for resolving the feedback from the starburst regions in its spiral arms that are spatially distinct from the AGN activity at the galactic center. 
A spatially unresolved molecular line survey by \citet{Aladro+2013} found this AGN-host galaxy being chemically distinct from pure starburst galaxies. 
The AGN is obscured by a molecular torus \citep{GB+2014}. 
Surrounding the torus is an elliptical region that is largely devoid of molecular gas, while this region is also characterised by strong emission from ionised gas in a biconal outflow driven by the AGN. 
Streams of gas seem to pass through this region, connecting the torus to the rest of the circumnuclear disk (CND). 
{The fueling of the AGN seems to be thwarted on an intermediate scale (r$<50$ pc) due to the outflowing nature of the CND and the torus-CND connections shown at comparable scales. }

The CND extends to $r\sim200$ pc, and its different regions show significantly distinct chemistry, in a manner that appears to match the spatial extent of the influence of UV and X-ray irradiation from the nucleus \citep{GB+2014}. 
All tracers observed by \citet{GB+2014} indicate a massive outflow from $r\sim50$pc out to $\sim 400$pc. 
This massive molecular outflow is likely launched by the interaction between the molecular gas in the CND and both the AGN ionized wind and the radio jet plasma, and is hence viewed as a sign of ongoing AGN feedback \citep{GB+2014,GB+2019}. 
This interaction has produced large-scale molecular shocks and the associated rich shock chemistry at different locations in the CND \citep{Viti+2014}. 
\citet{Viti+2014} found that models for the chemically distinct sub-regions of the CND each require a three-phase interstellar medium. {One component is indeed shocked gas, while the other two regions are with enhanced CRIR (and/or X-ray activities) by a least a factor of 10. This illustrates the complexity of the region. }
\citet{Scourfield+2020} reported that the temperature and the CRIR in the CND increase from east to west from observations of CS and subsequent modelling. 
Shocks in the CND of NGC 1068 were also studied with both observations and chemical modeling of HNCO and SiO \citep{Kelly+2017,Huang+2022,Huang_Viti_2023}, which are typically used as tracers of slow and fast shocks respectively. 

Gas phase methanol (\ce{CH3OH}) is commonly seen in the ISM, both in Galactic and extragalactic environments, and has been proposed as one of the shock tracer candidates. 
As the simplest complex organic molecule (COM), \ce{CH3OH} can be formed efficiently in cold environments (12-20K) through repeated hydrogenation of CO on ice grains \citep{Fuchs+2009} as well as through recently proposed H-abstraction reaction of radicals on ice grains \citep{Alvarez+2018, Simons+2020,Santos+2022}. 
A scheme for gas-phase formation has also been proposed, yet it remains incapable of accounting for the observed interstellar gas-phase \ce{CH3OH} \citep{Garrod+2006FaDi,Geppert+2006FaDi}. 
The \ce{CH3OH} formed on ice grains can then be released back to the gas phase via thermal sublimation or non-thermal desorption processes. 
Among these processes, shocks can effectively liberate the solid-state \ce{CH3OH} into gas phase through sublimation and sputtering. 
\ce{CH3OH} could, however, be destroyed in the presence of dissociative J shocks or fast shocks, as suggested in the observational work by \citet{Suutarinen+2014}. 
On the other hand, the chemical modeling work by \citet{Holdship+2017} has shown that non-dissociative high velocity C shocks (e.g. with shock velocity $>45$ km/s) can still elevate the gas-phase \ce{CH3OH} to comparable levels as achieved by the slow C shocks (e.g. with shock velocity $<20$ km/s. )

\ce{CH3OH} has two {spin isomers}\footnote{ {Here we refer to nuclear spin isomers. Molecules that have identical nuclei but nonzero nuclear spin could exist at different energy levels belonging to the so called “nuclear spin modifications” or “nuclear spin isomers" \citep{Hougen_Oka_2005}.}}, referred to as E- and A- \ce{CH3OH} owing to the combinations of nuclear spin alignment in the three H-atoms of the methyl (\ce{CH3}) group. 
In A-type \ce{CH3OH}, the nuclear spins of the three protons in the methyl group are parallel. 
For E-type \ce{CH3OH}, one of the protons in the methyl group has an anti-parallel nuclear spin with respect to the others. 
Observationally the two spin isomers of \ce{CH3OH} have different spectroscopic properties including transition frequencies. 

\ce{CH3OH} has been detected in multiple studies of the CND and the starburst ring (SB ring) of NGC 1068 at low-resolution ($>80$ pc) \citep{Aladro+2013,Takano+2014,Nakajima+2015}. 
In these results, however, only the \ce{CH3OH} $J_K = 2_{K} - 1_{K}$ transition has been detected. 
Observations of \ce{CH3OH} transitions covering a range of excitation energies will be critical in quantitatively characterize the gas properties traced by \ce{CH3OH}. 

In the current work, we present high resolution ($\sim 50$ pc scale) ALMA observations of the CND of NGC 1068 for six "clusters" of \ce{CH3OH} transitions with emission from both A- and E- \ce{CH3OH}. 
This high spatial resolution provides physical scales comparable to giant molecular clouds (GMCs). 
We analyzed these observations with non- local thermodynamic equilibrium (non-LTE), radiative transfer modeling. 
The goal is to quantitatively characterize the gas properties in order to gain insights into its association with the local shock history in the CND of NGC 1068. 
These observations can help better constrain the properties of the gas in the typical environments where COMs are expected to be found, and also explore the potential of using \ce{CH3OH} as a shock tracer. 
The paper is structured as follows. 
In Section 2, we describe the observations and the data reduction process. 
In Section 3, we present the molecular line intensity maps and the extracted velocity-integrated line intensities. 
In Section 4, we perform a non-LTE radiative transfer analysis in order to constrain the physical conditions of the gas. We discuss the potential association between the gas properties traced by \ce{CH3OH} and the presence of shocks. 
We also briefly discuss the E-/A- isomer ratio that is measured in this extragalactic source, and the potential indication of the \ce{CH3OH} formation environment. 
We briefly summarize our findings in Section 5. 
\section{Observations and data reduction}
\label{sec:obs}
\begin{table*}[t!]
  \centering
  \caption{List of \ce{CH3OH} transitions used in this study, ordered by rest frequencies. Note that only lines covered within $\pm 230$ km s\textsuperscript{-1} with respect to the main line\textsuperscript{(a)} are listed as shown with vertical lines in Figure \ref{fig:4p1_spec} - these are considered to be the "contributors" of the velocity-integrated intensities measured from the moment-0 maps. Observational details and the spatial resolution of the data used in this paper are also provided. A distance of 14 Mpc is assumed. }
  \label{tab:table_obsinfo}
  \begin{tabular}{c|cccccc}
  \hline
    Transition & Rest Frequency & E\textsubscript{u} & $g_u$ & $A_{ul}$ & Spatial resolution & mJy/beam to K  \\
    {} & [GHz] & [K] & {} & [s$^{-1}$] & {[$''$]} &{}\\
    \hline
    E- \ce{CH3OH} $J_K = 2_{-1} - 1_{-1}$ & 96.739358 & 12.5 & {5} & {$2.557794\times 10^{-6}$} & {$0.75\times0.62$} & {0.28} \\ 
    A- \ce{CH3OH} $J_K = 2_{0} - 1_0$ & 96.741371 & 7.0 & 5 & 3.408015$\times 10^{-6}$ &{- -} & {- -}\\
    E- \ce{CH3OH} $J_K = 2_{0} - 1_{0}$ & 96.744545 & 20.1 & {5} & $ 3.407341\times 10^{-6}$ & {- -} & {- -} \\
    E- \ce{CH3OH} $J_K = 2_{1} - 1_{1}$ & 96.755501 & 28.0 & {5} & $ 2.624407\times 10^{-6}$ & {- -} & {- -} \\
    \hline
    E- \ce{CH3OH} $J_K = 3_{0} - 2_{0}$ & 145.093754 & 27.1 & 7 & $1.231417\times 10^{-5}$ & {$0.49\times0.40$} & {0.30} \\
    E- \ce{CH3OH} $J_K = 3_{-1} - 2_{-1}$ & 145.097435 & 19.5 & 7 & $1.095695\times 10^{-5}$ & {- -} & {- -} \\
    A- \ce{CH3OH} $J_K = 3_{0} - 2_{0}$ & 145.103185 & 13.9 & 7 & 1.232344$\times 10^{-5}$ & {- -} & {- -} \\
    A- \ce{CH3OH} $J_K = 3_{-2} - 2_{-2}$ & 145.124332 & 51.6 & 7 & 6.892634$\times 10^{-6}$ & {- -} & {- -} \\
    E- \ce{CH3OH} $J_K = 3_{2} - 2_{2}$ & 145.126191 & 36.2 & 7 & $6.771773\times 10^{-6}$ & {- -} & {- -} \\
    E- \ce{CH3OH} $J_K = 3_{-2} - 2_{-2}$ & 145.126386 & 39.8 & 7 & $6.857968\times 10^{-6}$ & {- -} & {- -} \\
    E- \ce{CH3OH} $J_K = 3_{1} - 2_{1}$ & 145.131864 & 35.0 & 7 & $1.124576\times 10^{-5}$ & {- -} & {- -} \\
    A- \ce{CH3OH} $J_K = 3_{2} - 2_{2}$ & 145.133415 & 51.6 & 7 & $6.894658\times 10^{-6}$ & {- -} & {- -} \\
    \hline
    E- \ce{CH3OH} $J_K = 5_{0} - 5_{-1}$ & 157.178987 & 47.9 & 11 & $2.037818\times 10^{-5}$ & {$0.65\times0.41$} & {0.19} \\
    E- \ce{CH3OH} $J_K = 4_{0} - 4_{-1}$ & 157.246062 & 36.3 & 9 & $2.098481\times 10^{-5}$ & {- -} & {- -} \\
    E- \ce{CH3OH} $J_K = 1_{0} - 1_{-1}$ & 157.270832 & 15.4 & 3 & $2.205743\times 10^{-5}$ & {- -} & {- -} \\
    E- \ce{CH3OH} $J_K = 3_{0} - 3_{-1}$ & 157.272338 & 27.1 & 7 & $2.146271\times 10^{-5}$ & {- -} & {- -} \\
    E- \ce{CH3OH} $J_K = 2_{0} - 2_{-1}$ & 157.276019 & 20.1 & 5 & $2.181936\times 10^{-5}$ & {- -} & {- -} \\
    \hline
    A- \ce{CH3OH} $J_K = 1_{-1}-1_0$ & 303.366921 & 16.9 & 3 & 2.263319$\times 10^{-4}$ &{$0.46\times0.42$} & {0.07} \\
    A- \ce{CH3OH} $J_K = 2_{-1}-2_{0}$ & 304.208348 & 21.6 & 5 & 2.115671$\times 10^{-4}$ &{$0.46\times0.42$} & {0.07} \\
    A- \ce{CH3OH} $J_K = 3_{-1}-3_{0}$ & 305.473491 & 28.6 & 7 & 1.632485$\times 10^{-4}$ &{$0.48\times0.44$} & {0.06} \\
    \hline
    \hline
    \ce{c-/o-C3H2}  $J = 3_{1,2}-2_{2,1}$ & 145.0895946 & 16.05 & 21 & $6.764\times 10^{-5}$ &{} & {} \\
    SiO  $J = 7-6$ & 303.9268092 & 58.4 & 15 & $1.4648\times 10^{-3}$ &{} & {} \\
    OCS  $J = 25-24$ & 303.99326170 & 189.68 & 51 & $8.1948\times 10^{-5}$ &{} & {} \\
    SO  $J = 7_{8}-6_{7}$ & 304.0778440 & 62.1 & 17 & $3.609\times 10^{-4}$ &{} & {} \\
    \hline
  \end{tabular}\\
  \footnotesize{(a) The main line refers to transitions with the lowest $E_{u}$ in each frequency group - in particular for the 96 GHz, 145 GHz, and 157 GHz groups. }
\end{table*}
\begin{table}[t!]
  \centering
  \caption{Coordinates (RA and Dec) of the four selected regions within the CND. }
  \label{tab:table_7regs}
  \begin{tabular}{c|cc}
  \hline
    Name & RA & DEC   \\
    \hline
    CND R1  & 02\textsuperscript{h}42\textsuperscript{m}40\textsuperscript{s}.7617 & -00$^{\circ}$00${'}$48${''}$.1200 \\
    CND R2  & 02\textsuperscript{h}42\textsuperscript{m}40\textsuperscript{s}.7243 & -00$^{\circ}$00${'}$49${''}$.2400 \\
    CND R3  & 02\textsuperscript{h}42\textsuperscript{m}40\textsuperscript{s}.6030 & -00$^{\circ}$00${'}$48${''}$.9600 \\
    CND R4  & 02\textsuperscript{h}42\textsuperscript{m}40\textsuperscript{s}.6590 & -00$^{\circ}$00${'}$47${''}$.7000 \\
    \hline
  \end{tabular}
\end{table}

The \ce{CH3OH} transitions investigated in this paper were observed towards NGC 1068 using ALMA. 
These data were obtained during cycle 6 (project-ID: 2018.1.01506.S) with band 3 \& 4 receivers, and cycle 3 (project-ID: 2015.1.01144.S) using band 7 receivers. 
{The field of view (F.O.V.) is $56''$ for band-3 data, $42''$ for band-4 data, and $19''$  for band-7 data, respectively, which are all larger than the span of the CND ($\sim 5''$). 
In the band-3 and {band-4} observations, J0238+1636 was used as bandpass calibrator, and J0239-0234 as phase calibrator. 
In the band-7 observation, J0238+1636 was used as bandpass calibrator, and J0217+0144 as phase calibrator.
}
The above-mentioned data were calibrated and imaged using the ALMA reduction package CASA\footnote{http://casa.nrao.edu} \citep{CASA_2007}. 

The rest frequencies were defined using the systemic velocity determined by \citet{GB+2019}, as $v_{sys}$(LSR) = 1120 km s\textsuperscript{-1} (radio convention). 
The relative velocities throughout the paper refer to this $v_{sys}$. 
The phase tracking center was set to $\alpha_{2000}$ = (02\textsuperscript{h}42\textsuperscript{m}40.771\textsuperscript{s}), $\delta_{2000}$ = (–00$^{\circ}$00$'$47. 84$''$). 
The relevant information of each observation is listed in Table \ref{tab:table_obsinfo}. 
This table includes the target molecular transitions and the associated spectroscopic data, and the synthesized beam size for each observation. 
The beam sizes of our observations range between $\sim0\,.\!\!^{\prime\prime}5-0\,.\!\!^{\prime\prime}8$, or 35-56 pc in physical scales. 
This is comparable to the typical scale of giant molecular clouds (GMCs). 
The spectral resolution of the data cubes lies between $\sim 4-6$ km/s. 

\begin{figure}
        \centering
    \includegraphics[scale=0.48]{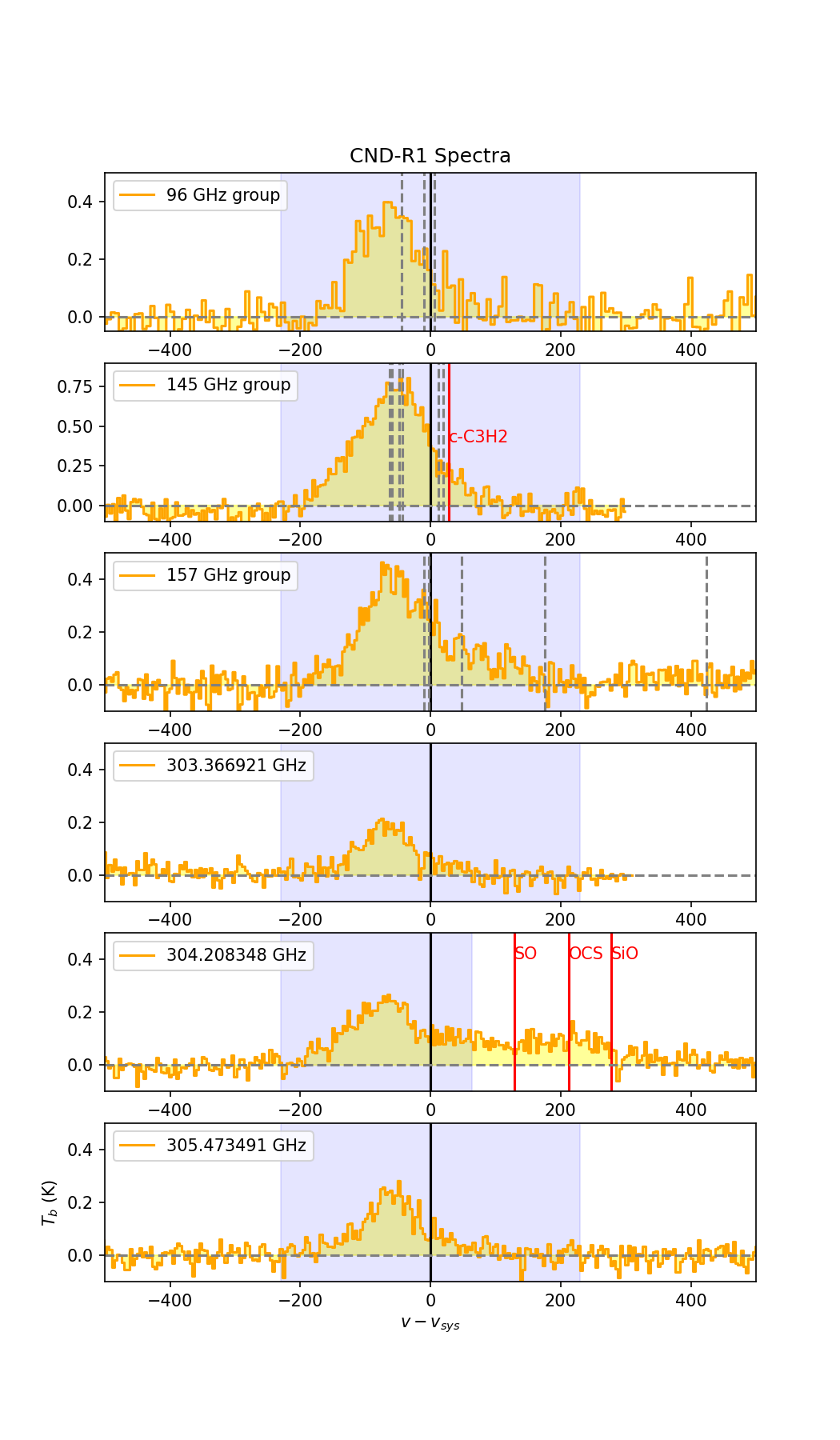}
    \caption{Spectra of all the \ce{CH3OH} transitions used in the current study. Each panel presents spectral data from the $0\,.\!\!^{\prime\prime}8\times0\,.\!\!^{\prime\prime}8$ CND-R1 region from the data cube at their original spectral resolution ($\Delta \varv \sim 4-6$ km/s) as solid orange curve. 
    The solid black vertical lines are the frequency references using \ce{CH3OH} lines with the lowest $E_{u}$ in each group. The grey dashed vertical lines are adjacent \ce{CH3OH} lines potentially covered by our spectral setup. Velocities refer to $v_{sys, LSRK}=1120$ km s\textsuperscript{-1}. The blue shaded area indicates the velocity coverage we use to derive the velocity-integrated line intensities in our analysis. The grey dashed horizontal line gives the 0-baseline reference. }
    \label{fig:4p1_spec}
\end{figure}
\section{Molecular line emission}
\subsection{Molecular-line spectra}
In order to investigate the physical structure traced by \ce{CH3OH} and to compare it with the shock study using \ce{SiO} and \ce{HNCO} observations at the same resolution by \citet{Huang+2022}, we followed their CND region selection: two regions (R1, R2) in the east of the CND, two regions (R3, R4) in the west of the CND.  Each region encompasses a beam size of  $0\,.\!\!^{\prime\prime}8\times0\,.\!\!^{\prime\prime}8$ to match the lowest angular resolution among our data and those from \citet{Huang+2022}. 
Table \ref{tab:table_7regs} lists these four selected positions within NGC 1068 with their coordinates, which are the center of the individual $0\,.\!\!^{\prime\prime}8\times0\,.\!\!^{\prime\prime}8$ apertures. 
Figure \ref{fig:4p1_spec} shows the sample spectra from the position CND-R1 (see Table \ref{tab:table_7regs}) for all transitions used in this study in units of [K]. We use spectra from CND-R1 as a representative sample as CND-R1 is the brightest spot in the CND in all transitions in the current work. 
All the [mJy beam\textsuperscript{-1}] to [K] conversion factors are listed in Table \ref{tab:table_obsinfo}. 

\subsubsection{Line blending between E- and A- isomers}
\label{sec:AEblending}
Figure \ref{fig:4p1_spec} also demonstrates the unavoidable line blending between A- and E- \ce{CH3OH} transitions in the line spectra (top three panels).  
The line width of \ce{CH3OH} is at least 50 km/s - based on the clean isolated \ce{CH3OH} $1_{-1}-1_{0}$ line at 303.366921 GHz) - and as the nearest line to that is only few km/s away (e.g top 3 panels of spectra) {we can not disentangle one from another by individual component fitting}. 
Given that the intensity contribution from each transition within the same blended group is a function of the spectroscopic properties of the group as well as the gas excitation conditions, generally one can not assume the transition with lowest $E_u$ to dominate the emission. 
We will revisit this in Section \ref{sec:radex}. 
\subsection{Moment-0 maps}
\label{sec:mom0}
\begin{figure*}
        \centering
    \includegraphics[scale=0.42]{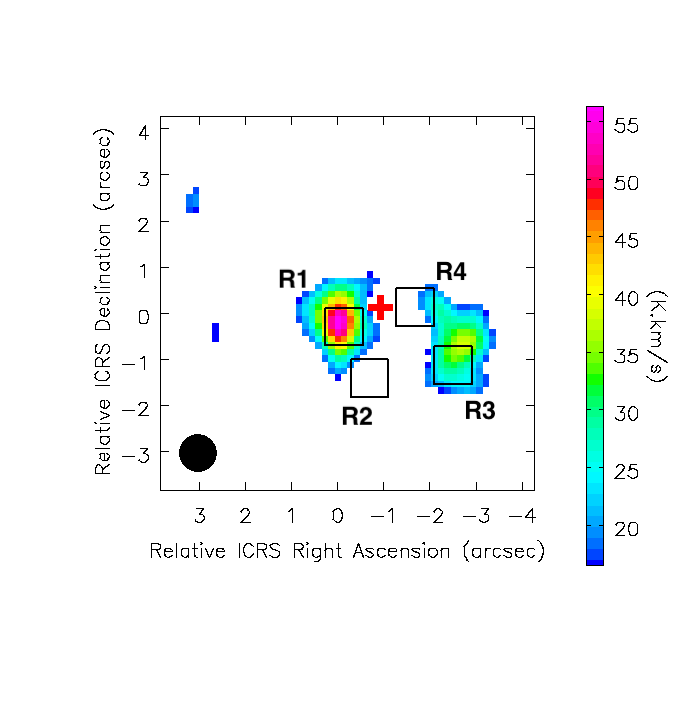}
    \includegraphics[scale=0.42]{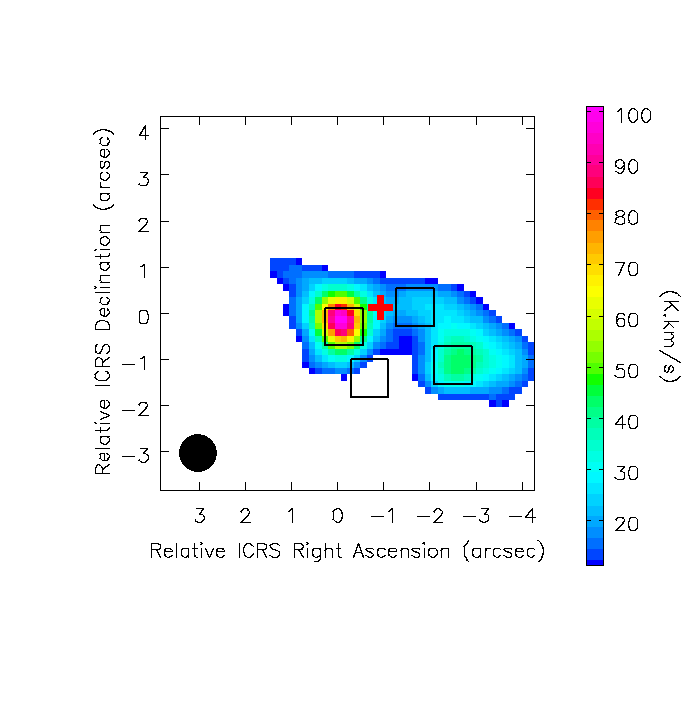}\\
    \includegraphics[scale=0.42]{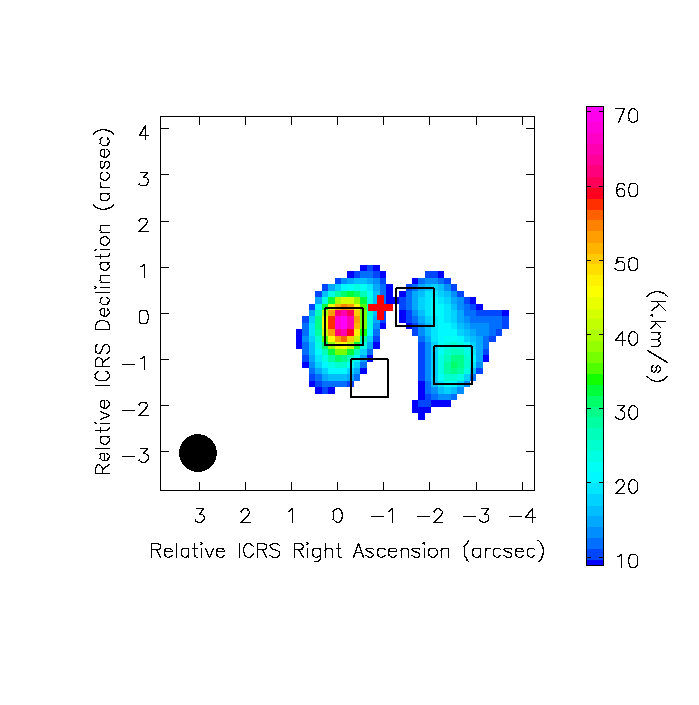}
    \includegraphics[scale=0.42]{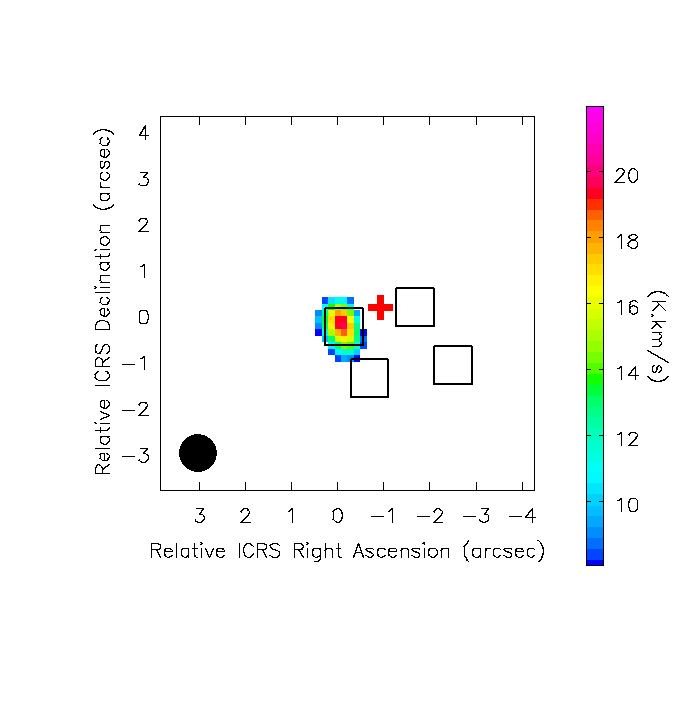}\\
    \includegraphics[scale=0.42]{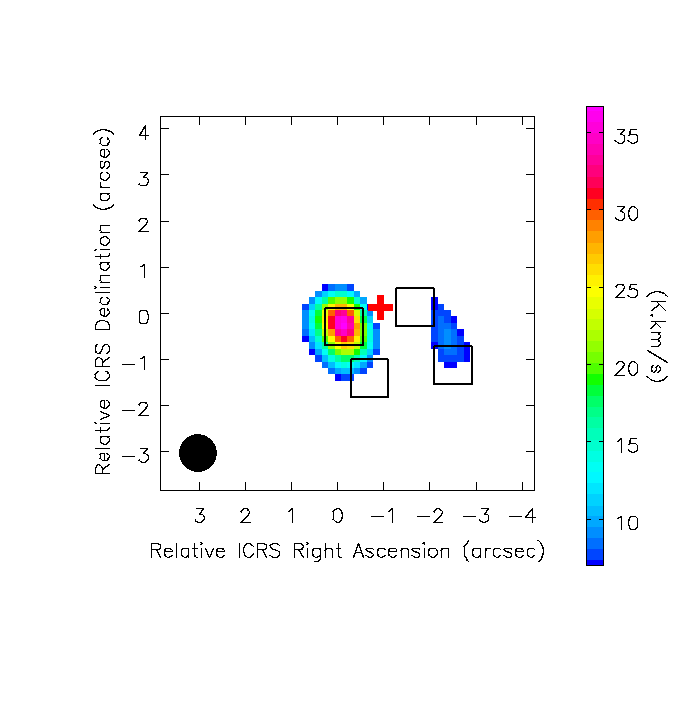}
    \includegraphics[scale=0.42]{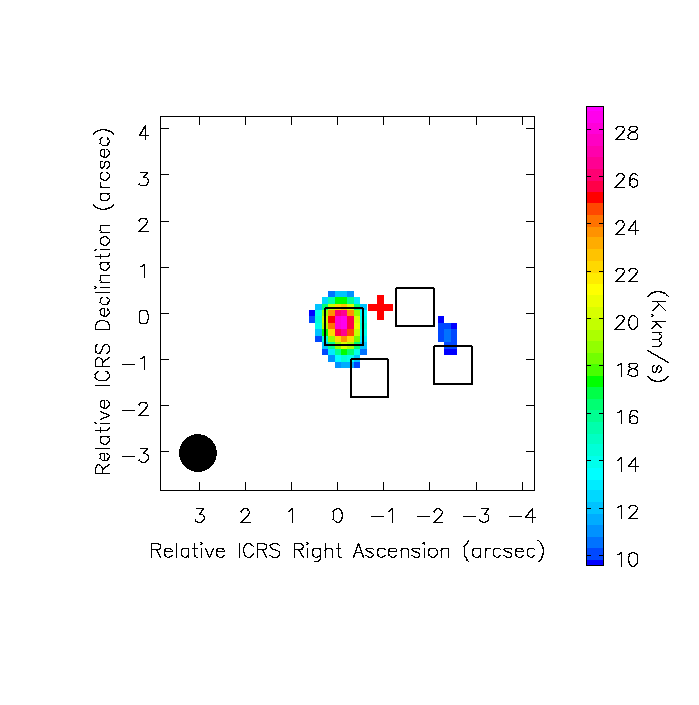}
    \caption{Moment-0 maps of all the \ce{CH3OH} transitions used in the current study. \textit{Top left:} 96 GHz group. \textit{Top right:} 145 GHz group. \textit{Middle left:} 157 GHz group. \textit{Middle right:} 303 GHz transition. \textit{Bottom left:} 304 GHz transition. \textit{Bottom right:} 305 GHz transition. The red cross on the map marks the AGN position derived by \citet{GB+2019}. }
    \label{fig:Maps_m3_mom0-I}
\end{figure*}
The moment-0 maps present the spatial distribution of line intensities integrated over velocity for each transition group as listed in Table \ref{tab:table_obsinfo}. 
All the moment-0 maps are subsequently smoothed to a common resolution of $0\,.\!\!^{\prime\prime}8$ and applied with a {$3.0 \sigma$} threshold clipping (Figure \ref{fig:Maps_m3_mom0-I}) 
The velocity span, over which the line fluxes were integrated, was set to include significant emission arising from velocity structures such as rotation and outflow motion in NGC 1068. 
In this work we used $|v-v_{sys}^{LSRK} |\leq$ 230 km s\textsuperscript{-1} to cover such velocity span. 
The only exception is for the A-\ce{CH3OH} $J_K = 2_{-1}-2_{0}$ line, which has {an} SO $J = 7_{8}-6_{7}$ line nearby, and the velocity integration was performed in a narrower range ($-130-64$ km/s) to not exceed the midpoint between the target \ce{CH3OH} line and the SO line. 
In the case of the \ce{CH3OH} line cluster at $\sim 145$ GHz, given the unavoidable blending with c-\ce{C3H2}, we use an existing estimate of the contribution of this particular c-\ce{C3H2} transition and the associated uncertainty based on observations towards another nearby galaxy, NGC 253 \citep{ALCHEMI_main_2021}, as an estimate for NGC 1068. 
{While we note that NGC 253 (a starburst) is very different from NGC 1068 (an AGN dominated galaxy), NGC 253 is the only galaxy where a quantitative estimate of the contribution from the c-\ce{C3H2} $J = 3_{1,2}-2_{2,1}$ transition to the methanol cluster has been performed. This estimate is derived from multi-transitional line fitting using data from ALMA band 3 to band 7 and the line fitting program \texttt{MADCUBA} \citep{ALCHEMI_main_2021}. }
Table \ref{tab:table_integ_I} lists all the measured velocity-integrated line intensities (in K km/s) from the observations with common resolution of $0\,.\!\!^{\prime\prime}8$. 

\section{Inference of gas properties in the CND}
\label{sec:gas_properties}
The configuration of the CND in NGC 1068 relative to its jet-ISM interaction and the associated outflow geometry makes it a prime source to investigate shock-driven chemistry \citep{GB+2014,Viti+2014}. 
The axes of the ionized outflow and the jet both form a small angle relative to the molecular gas disk of the galaxy. This particular geometry implies that there is a strong coupling between the AGN outflow and the jet with the ISM in the disk out to r$\sim400-500$ pc. 
Within this context we wish to characterize the gas properties traced by \ce{CH3OH} across the CND regions, and to study its association with the shock history of the CND regions. 

In characterizing the gas properties traced by any molecular species with multiple transitions, one often assumes optically thin Local Thermal Equilibrium (LTE) conditions. {This allows the implementation of the so-called "rotation diagram" \citep{Goldsmith_Langer_1999} based on the observed line intensities from individual molecular transition.} 
One can then infer the physical properties such as rotational temperature and column density of {the species from this rotation diagram}. 
Such approach relies not only on the validity of the assumption of optically thin LTE, but also requires each transition to be free of contamination from other lines. 
In our case, however, blending from different transitions is prevalent (both from A- and E- \ce{CH3OH}) and {hard to disentangle by individual component fitting as mentioned in Sect. \ref{sec:AEblending}.} One option is to assume thermalization of the gas and to re-distribute the measured line intensity to all transitions involved in the line blending \citep{Martin+2006}. 

Alternatively, we can use non-LTE radiative transfer modeling and infer the gas properties {by comparing the summation of the predicted line intensities from all blending lines in each "transition group" with the observed line intensity for each specific measurement}. 
To characterize the \ce{CH3OH} emissions across the CND region, we choose indeed to perform a radiative analysis via non-LTE radiative transfer modeling. 
With this approach we are able to quantitatively characterize the column density of \ce{CH3OH} and, more importantly, the gas temperature and the gas density in the CND. 
\label{sec:ladder}
\begin{table*}[ht!]
  \centering
  \caption{Velocity-integrated line intensity of \ce{CH3OH} transitions covered in the current work. These values are extracted from data that have been smoothed to the common $0\,.\!\!^{\prime\prime}8$ resolution, as indicated in Section \ref{sec:mom0}. }
  \label{tab:table_integ_I}
  \begin{tabular}{c|cccc}
  \hline
    CND region & R1 & R2 & R3 &R4 \\
    Transition  & I($\sigma$) & I($\sigma$)  & I($\sigma$) &  I($\sigma$) \\
    {}  & [K km/s] & [K km/s] & [K km/s] & [K km/s]\\
    \hline
    \hline
    {96 GHz group} &  46.104(7.002)  & 5.253(1.353)  & 29.987(4.630)  &  13.005(2.239)  \\
    {145 GHz group} &  41.053(41.059)  &  2.995(3.085)  & 19.891(19.904) &  11.043(11.068)  \\
    {157 GHz group} &  58.144(8.742)  &  4.125(0.856)  & 25.645(3.892)  & 13.975(2.178)  \\
    {A- \ce{CH3OH} $J_K = 1_{-1}-1_0$} &  15.987(2.457)  &  2.125(0.624)  & 5.202(0.947)  & 2.918(0.692)  \\
    {A- \ce{CH3OH} $J_K = 2_{-1}-2_{0}$} &  30.962(4.668)  &  5.099(0.896)  & 6.904(1.136)  & 4.671(0.841)  \\
    {A- \ce{CH3OH} $J_K = 3_{-1}-3_{0}$} &  23.871(3.637)  &   4.033(0.879)  & 7.203(1.255)  & 5.532(1.047)\\
    \hline
  \end{tabular}
\end{table*}
\label{sec:radex}
For the non-LTE analysis, we use the radiative transfer code \texttt{RADEX} \citep{radex_vandertak_2007} via the Python package SpectralRadex\footnote{https://spectralradex.readthedocs.io} \citep{Holdship+2021_SpectralRadex}. 
The molecular data used for \ce{CH3OH} are from \citet{CH3OH_mol_data_2010} taken from the LAMDA database \citep{LAMDA_2005}. 
We coupled the RADEX modeling with the Markov Chain Monte Carlo (MCMC) sampler \texttt{emcee}  \citep{FM+2013_emcee} to perform a Bayesian inference of the parameter probability distributions in order to properly sample this parameter space and obtain reliable uncertainties. 
We assume priors of uniform or log-uniform distribution within the determined ranges (listed in Table \ref{tab:table_prior}). 
With this, we assume that the uncertainty on our measured intensities is Gaussian so that our likelihood is given by $P(\theta | d) \sim \exp(-\frac{1}{2}\chi^2),$ where $\chi^2$ is the chi-squared statistic between our measured intensities and the \texttt{RADEX} predictions based on a set of parameters $\theta$. 
This allows us to assess the influence of physical conditions, such as gas density and temperature, in the excitation of the transitions analyzed. 

Regarding the blending of adjacent A- and E- isomer transitions, we can not distinguish them   and contributions from any of these lines is generally not negligible for our observations. 
In other words, all components in the same blended group should be taken into consideration in particular when one tries to compare the line intensity with radiative transfer modeling. 
Hence, we need to determine a priori the flux contribution from both A- and E- \ce{CH3OH} transitions in our \texttt{RADEX} modeling. 
The species column density from both isomers, $N_{A-}$ and $N_{E-}$, are, therefore, also explored as free parameters. 
In summary, the following physical parameters are explored: the gas volume density ($n_{H2})$, gas kinetic temperature ($T_{kin}$), the species column density ($N_{E-}$ and $N_{A-}$), and the beam filling factor ($\eta_{ff}$). 
The prior ranges and distributions are given in Table \ref{tab:table_prior}. 
We list the best fit for all the explored parameters in Table \ref{tab:Bayesian_gasphys}. 

Generally the gas densities and the molecular column densities of both A- and E- isomers are well constrained. 
The inferred A- and E- \ce{CH3OH} column densities are both between $10^{15-16}$ cm\textsuperscript{-2}. 
The physical interpretation regarding these measured A- and E- \ce{CH3OH} column densities will be further explored in Section \ref{sec:EAratio}. 
The gas densities point to $\sim 10^{6}$ cm\textsuperscript{-3} across all the CND regions. 
{This is not surprising since, as discussed in the Introduction, \ce{CH3OH} is a product of surface reactions in dense environments.  }
The gas kinetic temperature, however, cannot be well constrained in any of the CND regions. 
Despite the fact that the number of transitions involved in our analysis is greater than the number of free parameters we explore here, it is possible that the coverage of the excitation energy levels is not wide enough to fully constrain the gas temperature. 
We speculate that \ce{CH3OH} transitions of higher $E_{u}$ may be needed to properly constrain gas temperature. 
We note that we also tried a different algorithm to perform the Bayesian inference - a nested sampling Monte Carlo algorithm MLFriends \citep{ultranest16,ultranest19} using the \texttt{UltraNest} package\footnote{\url{https://johannesbuchner.github.io/UltraNest/}} \citep{ultranest21} - complementing our analysis with the Markov chain Monte Carlo (MCMC) sampler \texttt{emcee} \citep{FM+2013_emcee}. 
The inference results with \texttt{UltraNest} still failed to constrain the gas temperature.
\subsection{Are the gas properties consistent with shocks?}
As shown in Table \ref{tab:Bayesian_gasphys}, all the inferred gas densities point to $\sim 10^{6}$ cm\textsuperscript{-3} across all the CND regions. 
With such high gas density the post-shock gas is expected to cool off fairly quickly \citep[$\leq$ few tens of years,][]{Huang_Viti_2023}, which is somewhat consistent with the indication that the (poorly constrained) inferred temperatures are likely to be low ($\leq$ 100 K; see Figure 3). 

Based on the chemical modeling with \texttt{UCLCHEM} for the physical condition of the CND of NGC 1068 performed by \citet{Huang_Viti_2023}, both fast and slow shocks can enhance the methanol fractional abundance to comparable levels in the presence of shocks. 
However, in the chemical models without shocks, a higher gas temperature is required (such that water ice sublimates, $T\gtrsim 100$ K) to boost the \ce{CH3OH} abundance to comparable levels \citep{Huang_Viti_2023}. 
In models where a shock is occurring, the kinetic temperature of the gas can be very high due to shock heating but can also be low due to rapid post shock cooling in a dense medium. Both scenarios lead to methanol abundances compatible with observations. 
The gas temperature measured from observations, therefore, is one of the key physical parameters in discerning the chemical origin of the observed \ce{CH3OH}. 
Unfortunately in all four CND regions explored, none of the gas temperatures are constrained  enough to discern among shock scenarios. Nevertheless, qualitatively, most regions (CND-R1, CND-R3, and CND-R4) tend to point to $T_{kin}<100$ K. 
This would point to the shock origin of the observed gas-phase \ce{CH3OH} for in low temperature regimes only shocks can account for the enhanced gas-phase \ce{CH3OH} \citep{Huang_Viti_2023}. 

\subsection{Comparison with other shock tracers - SiO and HNCO}
We choose to compare our results with the work by \citet{Huang+2022} where SiO and HNCO maps of the same region and at the same spatial resolution are presented.  
As far as the spatial distribution of \ce{CH3OH} is concerned, it is more concentrated towards the east of the CND and much less prominent in the west of the CND. This is  similar to the SiO emission reported by \citet{Huang+2022}.
In contrast, the HNCO reported by \citet{Huang+2022}, even in the higher-J level (HNCO 6-5) is much more extended throughout the CND. 
Interestingly, in terms of the gas density, \ce{CH3OH} points to a higher gas density than HNCO, while SiO mostly points to lower gas densities near $10^3$ cm\textsuperscript{-3}. 
Finally, the gas temperature probed by \ce{CH3OH}, although not well constrained, is in the same range as that probed by HNCO. 
Compared to the high-$T$ gas traced by SiO, the lower temperature probed by HNCO and \ce{CH3OH} may be a product of faster cooling in denser gas environments regardless of the speed of the shock. 

From this comparison it remains inconclusive whether \ce{CH3OH} exclusively traces slow and non-dissociative shocks ($v_{s}<20$ km/s) as proposed by \citet{Suutarinen+2014} or whether it traces both fast- and slow- shock environments as hinted by the chemical modeling in \citet{Holdship+2017} and \citet{Huang_Viti_2023}. 

\subsection{E-/A- \ce{CH3OH} Ratio}
\label{sec:EAratio}
Aside from the gas density and gas temperature, the inferred $N_{E-}$ and $N_{A-}$ will be interesting to inspect. 
By assuming A- and E- \ce{CH3OH} isomers are formed and well mixed in the same environment, the equilibrium E/A ratio is linked to the temperature of the formation environment \citep{Wirstrom+2011,Holdship+2019}. 
Generally the conversion of one \ce{CH3OH} isomer to another is considered a strongly forbidden process; the abundance ratio of the two isomers is therefore often assumed to be fixed at the time of the \ce{CH3OH} formation \citep{Kawakita_Kobayashi_2009,Wirstrom+2011}. 
In particular, \citet{Wirstrom+2011} noted that \ce{CH3OH} is expected to be destroyed fairly rapidly compared to the time needed to alter the ratio. 
This implies that this ratio is thermalized at its formation and hence can provide insights into the origin and physical conditions of the methanol formation \citep{Wirstrom+2011}. 
In other words, the measured E/A ratio actually informs us about the local temperature associated with \ce{CH3OH} formation \footnote{Note that this formation temperature is set in the formation history of \ce{CH3OH}, and is different from the "present-time" gas temperature inferred at the beginning of Section \ref{sec:radex}} at different locations in the CND. 

On the ices, the optimal temperature for successive hydrogenation of CO to occur for methanol formation is suggested to be around $10-15$ K with a CO:\ce{H2O} ice mixture or pure solid-phase CO \citep{Watanabe+2004}. 
At this low-temperature regime, an overabundance of A-\ce{CH3OH} is expected ($N_{E-}/N_{A-} <1.0$). 
In our measured $N_{E-}/N_{A-}$ ratio, the most probable ratio values are fairly close to unity (see Table \ref{tab:Bayesian_gasphys}). 
This is interestingly different from the generally expected overabundance of A-\ce{CH3OH} \citep{Wirstrom+2011} and the observed $N_{E-}/N_{A-}$ ratio from Galactic outflows reported by \citet{Holdship+2019}. 
{The Galactic measurements of $N_{E-}/N_{A-}$ ratio reported by \citet{Holdship+2019} are all below 0.7 and point to methanol formation temperatures below 10 K (see Figure 4 in \citet{Holdship+2019}). 
In contrast, our measured ratio, which is close to unity, points to a higher formation temperature of \ce{CH3OH},  $\gtrsim 15$ K. Note that the uncertainty derived from the inferred column densities of both isomers remain fairly large, so this is still highly uncertain. }

{We also want to highlight the only other two existing E-/A- ratio measurement. 
Similar to our non-LTE modeling approach, \citet{Muller+2021} performed non-LTE modeling for A- and E- \ce{CH3OH} isomer observations of the z=0.89 molecular absorber toward PKS 1830-211. They presented a comparably high E-/A- ratio of $\sim1.0 \pm 0.1$ via the submillimeter absorption lines of methanol from this source \citep{Muller+2021}. 
There is also an LTE model for the nearby starburst galaxy NGC 253 \citep{Humire+2022}. They report a fairly wide range of $N_{E-}/N_{A-}$ ratio varying from $\sim1.1-5.0$ \citep[Table 3,][]{Humire+2022}. }
Since the existing gas-phase proton exchange described in the literature {is} considered dissociative \citep{Wirstrom+2011}, such {Galactic-extragalactic distinction} in $N_{E-}/N_{A-}$ ratio could imply a distinction between the formation environment of methanol in the nucleus of NGC 1068 from that of our own galaxy. 
Otherwise non-dissociative proton exchange reaction(s) such as proton exchange in the solid phase needs to be considered if one assumes that the $N_{E-}/N_{A-}$ ratio in Galactic and extragalactic environments start at a similar level. 
\begin{table}[ht!]
  \centering
  \caption{Prior range adopted for our parameter space explored in the \texttt{RADEX}-Bayesian inference process described in Sect. \ref{sec:radex}. The beam filling factor is defined as:  $\eta_{ff}=\frac{\theta^{2}_{S}}{\theta^{2}_{MB}+\theta^{2}_{S}}$}
  \label{tab:table_prior}
  \begin{tabular}{c|cc}
  \hline
    Variable  & Range & Distribution type\\
    \hline
    Gas density, $n_{\rm H2}$ [cm\textsuperscript{-3}] & $10^{2}-10^{8}$ & Log-uniform\\
    Gas temperature, $T_{\rm kin}$ [K] & $10-800$ & Uniform\\
    $N$(A-\ce{CH3OH}) [cm\textsuperscript{-2}] & $10^{12}-10^{18}$ & Log-uniform\\
    $N$(E-\ce{CH3OH}) [cm\textsuperscript{-2}] & $10^{12}-10^{18}$ & Log-uniform\\
    Beam filling factor, $\eta_{ff}$ & $0.0-1.0$ & Uniform\\
    \hline
  \end{tabular}
\end{table}
\begin{table*}
    \centering
    \caption{Inferred gas properties traced by \ce{CH3OH} from the Bayesian inference processes over four selected regions across CND (R1-R5). For poorly constrained cases, we identify the upper or lower limit of the distribution and place the percentile values of 95 (for upper limit) or 5 (for lower limit) in parenthesis. }
    \label{tab:Bayesian_gasphys}
    \begin{tabular}{cccccc}
        \hline
		Model & log(n[H2]) & Tkin & log(N[A-CH3OH]) & log(N[E-CH3OH]) & Filling factor \\ 
		\hline
		CND-R1 & $6.25^{+0.33}_{-0.27}$ & {- -} & $15.73^{+0.45}_{-0.29}$ & $15.89^{+0.47}_{-0.30}$ & {- -} \\ 
		CND-R2 & $6.49^{+0.96}_{-0.38}$ & $66^{+514}_{-35}$ & $15.00^{+0.55}_{-0.32}$ & $14.82^{+0.50}_{-0.37}$ & {- -} \\ 
		CND-R3 & $5.87^{+0.28}_{-0.26}$ & {- -} & $15.19^{+0.64}_{-0.42}$ & $15.49^{+0.58}_{-0.54}$ & {- -} \\ 
		CND-R4 & $5.97^{+0.35}_{-0.28}$ & {- -} & $15.04^{+0.54}_{-0.25}$ & $15.25^{+0.58}_{-0.29}$ & {- -} \\ 
		\hline
    \end{tabular}
\end{table*}

\begin{figure*}
  \sidecaption
  \includegraphics[width=18cm]{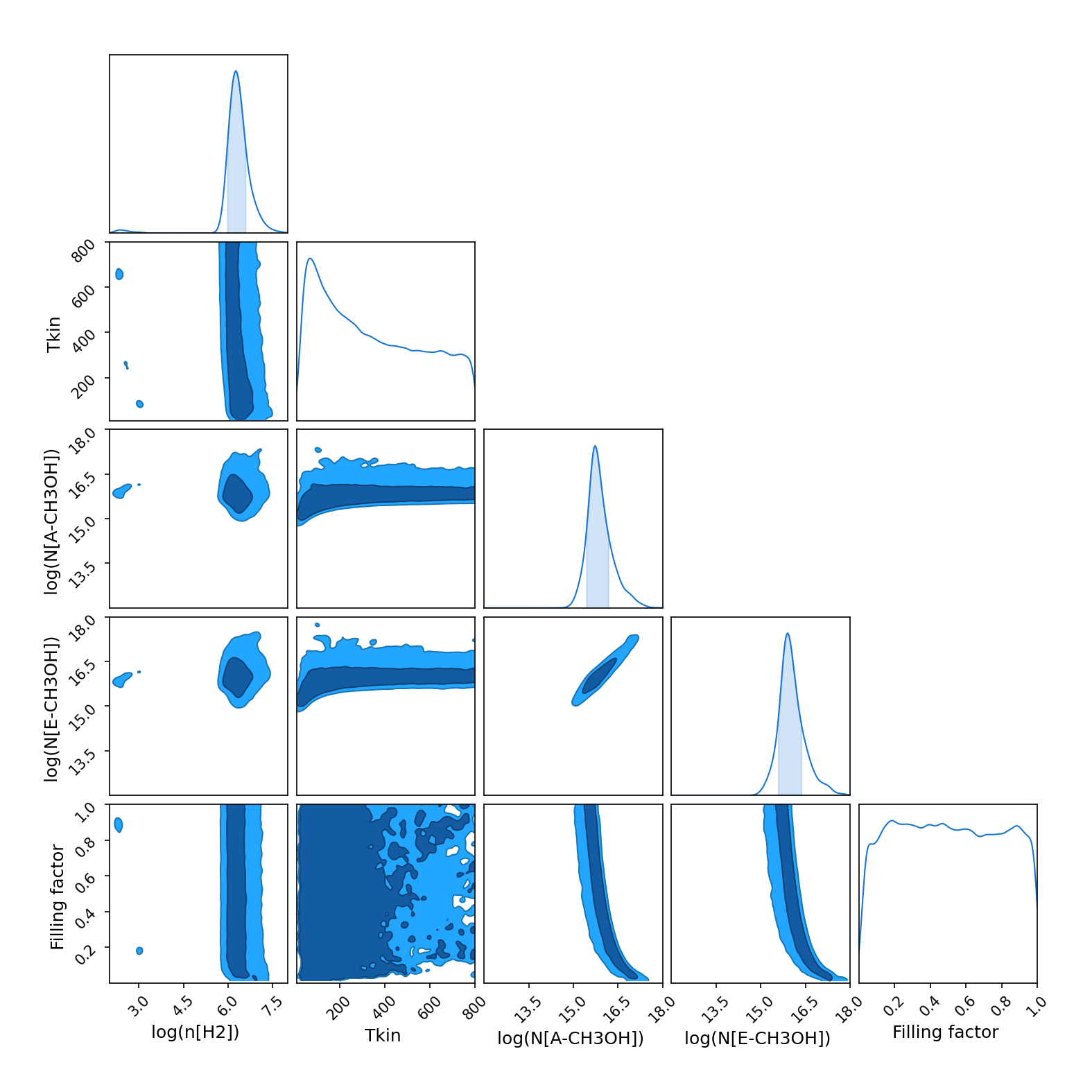}
  \caption{Bayesian inference results for gas properties traced by \ce{CH3OH} of the CND-R1 region. The corner plots show the sampled distributions for each parameter, as displayed on the x-axis. The 1-D distributions on the diagonal are the posterior distributions for each explored parameter;the reset 2-D distributions are the joint posterior for the corresponding parameter pair on the x- and y- axes. In the 1-D distributions, the $1\sigma$ regions are shaded with blue; both $1\sigma$ and $2\sigma$ are shaded in the 2-D distributions. }
  \label{fig:Baye_Overlay_R1}
\end{figure*}
\begin{figure*}
  \centering
  \begin{tabular}[b]{@{}p{1.0\textwidth}@{}}
    \centering\includegraphics[width=1.0\linewidth]{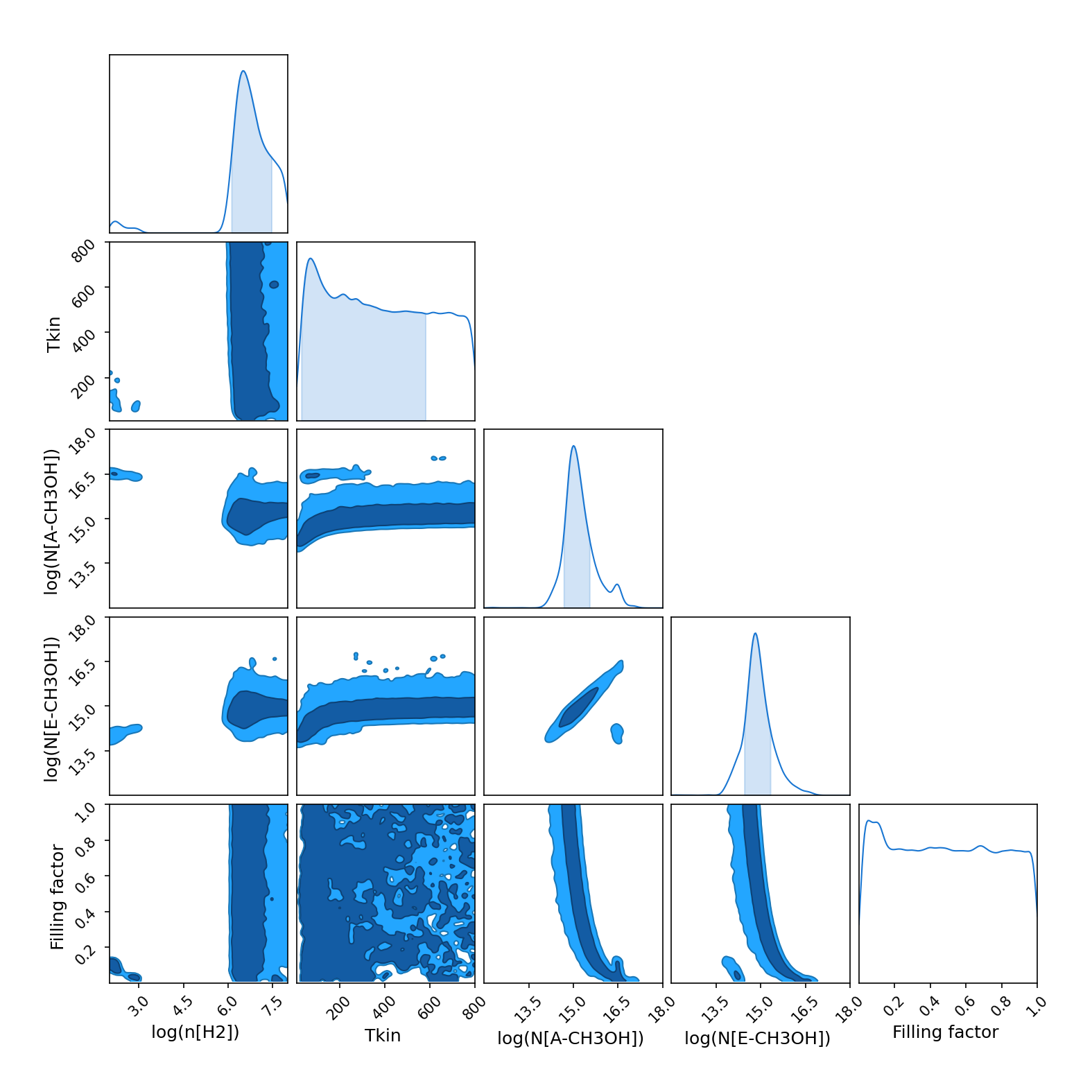} 
  \end{tabular}%
  \caption{As in Figure~\ref{fig:Baye_Overlay_R1} but for the CND-R2 region. This shows the Bayesian inference results for gas properties traced by \ce{CH3OH} of CND-R2 region. }
  \label{fig:Baye_Overlay_R2}
\end{figure*}
\begin{figure*}
  \centering
  \begin{tabular}[b]{@{}p{1.0\textwidth}@{}}
    \centering\includegraphics[width=1.0\linewidth]{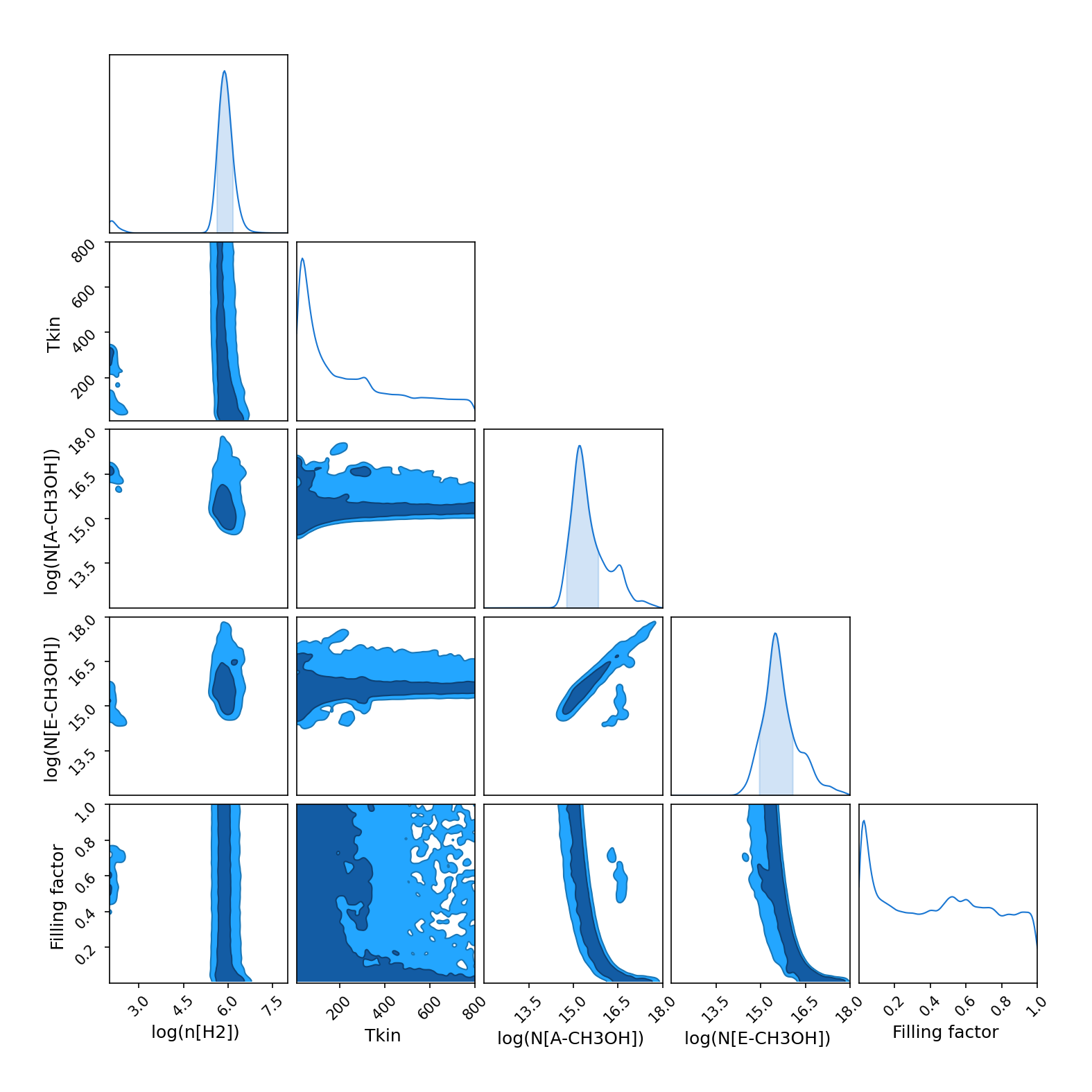} 
  \end{tabular}%
  \caption{ As in Figure~\ref{fig:Baye_Overlay_R1} but for the CND-R3 region. This shows the Bayesian inference results for gas properties traced by \ce{CH3OH} of CND-R3 region. }
  \label{fig:Baye_Overlay_R3}
\end{figure*}
\begin{figure*}
  \centering
  \begin{tabular}[b]{@{}p{1.0\textwidth}@{}}
    \centering\includegraphics[width=1.0\linewidth]{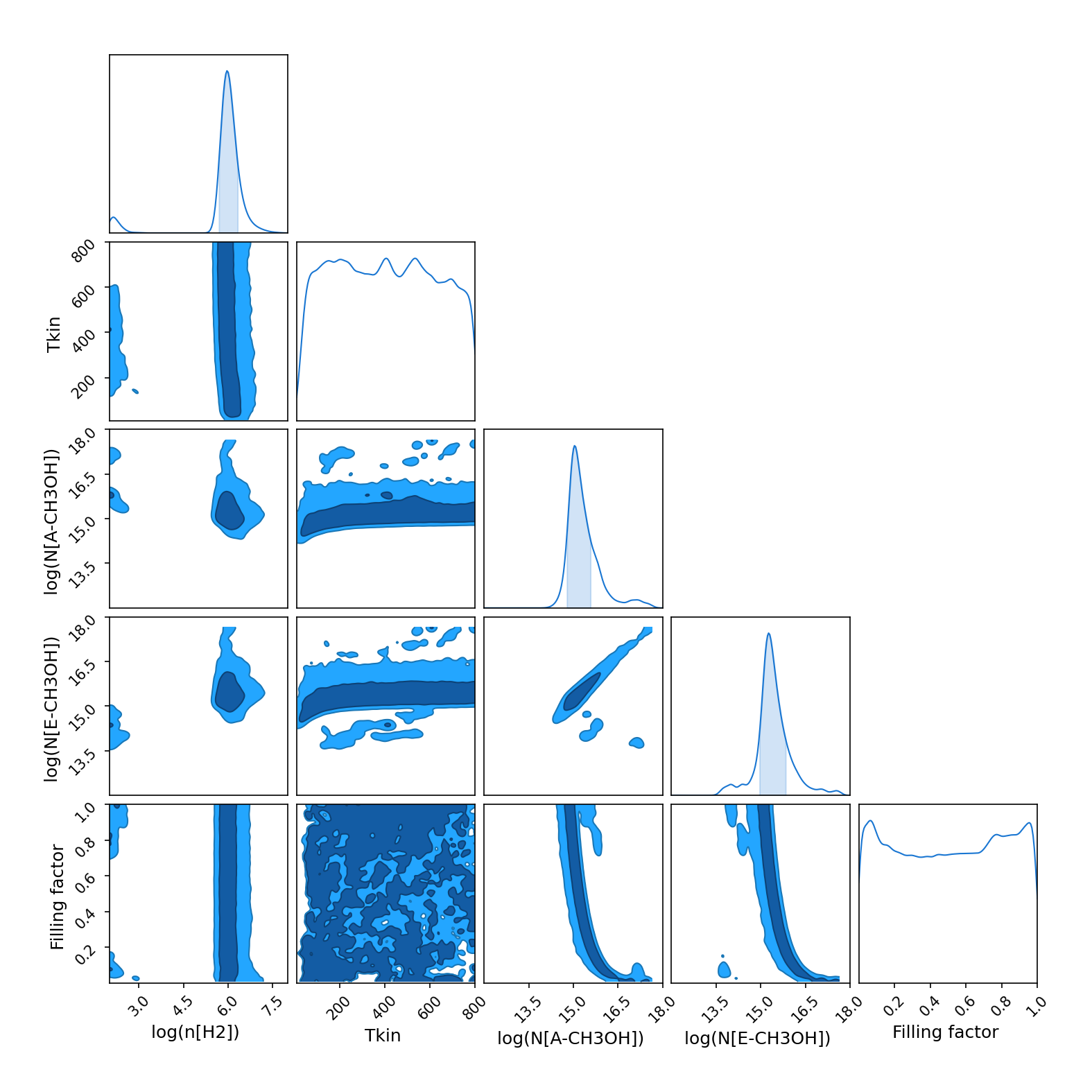} 
  \end{tabular}%
  \caption{As in Figure~\ref{fig:Baye_Overlay_R1} but for the  CND-R4 region. This shows the Bayesian inference results for gas properties traced by \ce{CH3OH} of CND-R4 region. }
  \label{fig:Baye_Overlay_R4}
\end{figure*}
\renewcommand{\arraystretch}{1.5}

\section{Conclusions}
The outflowing molecular gas in the CND of NGC 1068 and the associated large spread of velocities ($\sim100$ km s\textsuperscript{-1}) likely drive a variety of molecular shocks at different locations in the CND. 
We perform a multi-line molecular study with ALMA of \ce{CH3OH} in order to determine the chemical signatures of such molecular shocks in the CND of NGC 1068 at spatial scale of GMC ($0\,.\!\!^{\prime\prime}8\sim 50$ pc). 
We briefly summarize below our conclusions: 
   \begin{enumerate}
      \item We performed non-LTE radiative transfer analyses where we coupled RADEX with a Bayesian inference procedure, in order to infer the gas properties traced by \ce{CH3OH}. 
      \item The gas densities traced by \ce{CH3OH} point to $\sim 10^{6}$ cm\textsuperscript{-3} across all the CND regions. The gas kinetic temperature cannot be well constrained in any of the CND regions though the inferred temperature is likely to be low ($\lesssim 100$ K), suggesting a shock origin of the observed gas-phase \ce{CH3OH}. 
      \item We compared both spatial distribution of emissions from multiple tracers of different types of shocks (SiO and HNCO) versus \ce{CH3OH}, and the gas properties traced by these three species. It remains inconclusive, however, whether \ce{CH3OH} exclusively trace slow shocks or whether it traces both fast- and slow- shocks. 
      \item We also note that the inferred E-/A- isomer ratio is fairly close to unity, which is interestingly different from the Galactic measurements in the literature. 
   \end{enumerate}
\begin{acknowledgements}
      KYH, and SV received funding from the European Research Council (ERC) Advanced Grant MOPPEX 833460.vii. 
      SGB acknowledges support from research project grants PID2019-106027GA-C44 and PID2022-138560NB-I00 of the Spanish Ministerio de Ciencia e Innovaci{\'o}n. 
      KYH acknowledges assistance from Allegro, the European ALMA Regional Center node in the Netherlands. 
      This paper makes use of the following ALMA data: ADS/JAO.ALMA\#2013.1.00221.S, ADS/JAO.ALMA\#2015.1.01144.S, and ADS/JAO.ALMA\#2018.1.01506.S. ALMA is a partnership of ESO (represent- ing its member states), NSF (USA) and NINS (Japan), together with NRC (Canada), MOST and ASIAA (Taiwan), and KASI (Republic of Korea), in co-operation with the Republic of Chile. The Joint ALMA Observatory is operated by ESO, AUI/NRAO and NAOJ.
\end{acknowledgements}
\bibliographystyle{aa}
\bibliography{NGC1068,fundamentals,Additional}
\newpage
\begin{appendix}
\section{Comparison of the predicted intensity from RADEX with observed values: A posterior predictive check (PPC)}
In this section, we perform a posterior predictive check \citep[PPC,][]{Gelman+1996} for the inferred gas properties in Section \ref{sec:radex} using RADEX and Bayesian inference process. 
We do this by performing a comparison of the distribution of the observational data, the measured velocity-integrated intensities, to the predicted distribution described by our Bayesian model. 
This is to verify our posterior distribution produces a distribution for the data that is consistent with the actual data. 
We sample the predicted line intensities from our {posteriors} between 16-84 percentiles, and plot these against the observed line intensities. 

\begin{figure*}
  \centering
  \begin{tabular}[b]{@{}p{0.45\textwidth}@{}}
    \centering\includegraphics[width=1.0\linewidth]{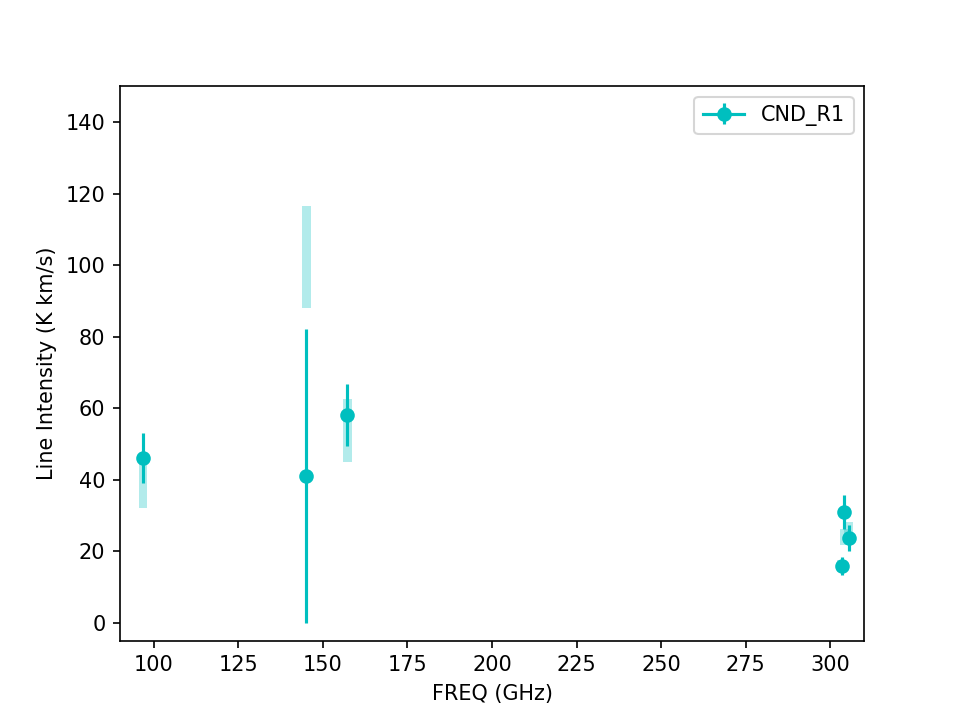} \\
    \centering\small (a) Observed and PPC intensities of \ce{CH3OH} in CND-R1
  \end{tabular}%
  \quad
  \begin{tabular}[b]{@{}p{0.45\textwidth}@{}}
    \centering\includegraphics[width=1.0\linewidth]{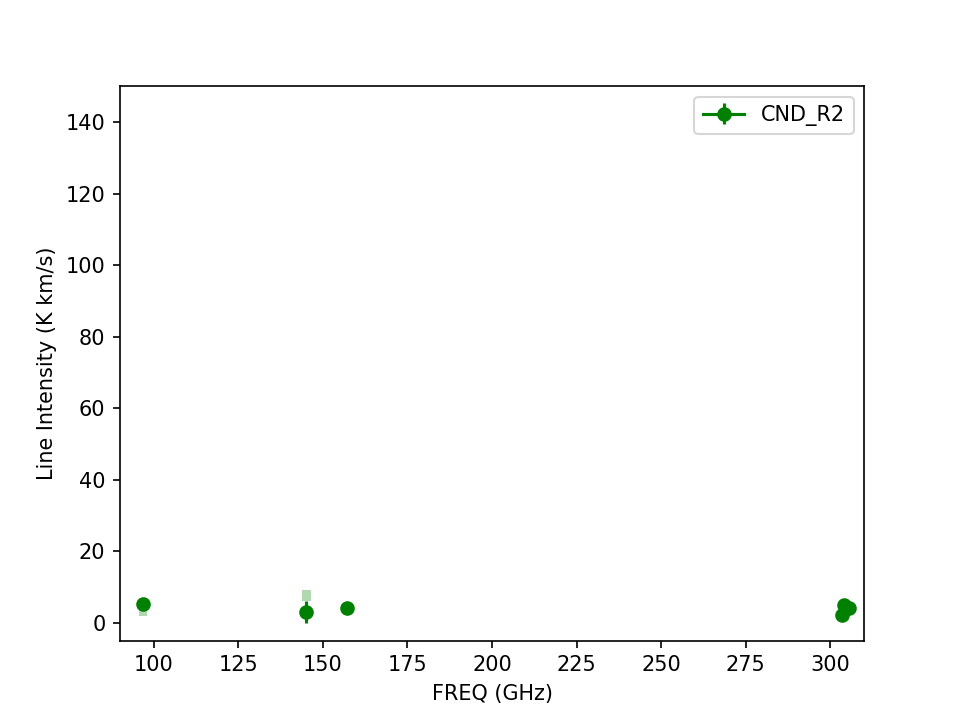} \\
    \centering\small (b) Observed and PPC intensities of \ce{CH3OH} in CND-R2
  \end{tabular}
  \begin{tabular}[b]{@{}p{0.45\textwidth}@{}}
    \centering\includegraphics[width=1.0\linewidth]{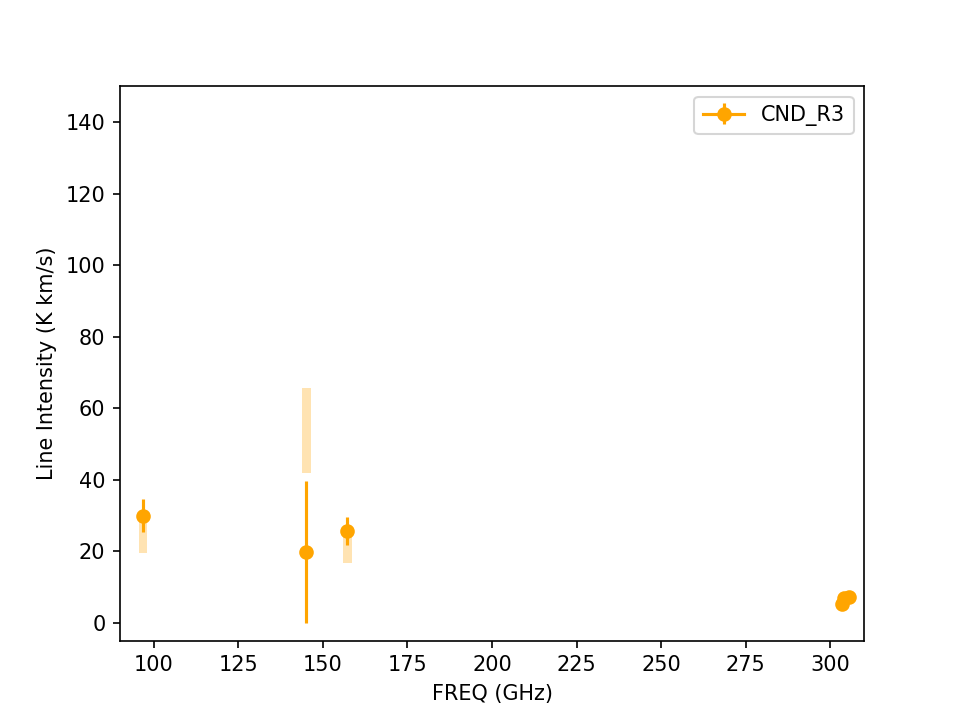} \\
    \centering\small (c) Observed and PPC intensities of \ce{CH3OH} in CND-R3
  \end{tabular}
  \begin{tabular}[b]{@{}p{0.45\textwidth}@{}}
    \centering\includegraphics[width=1.0\linewidth]{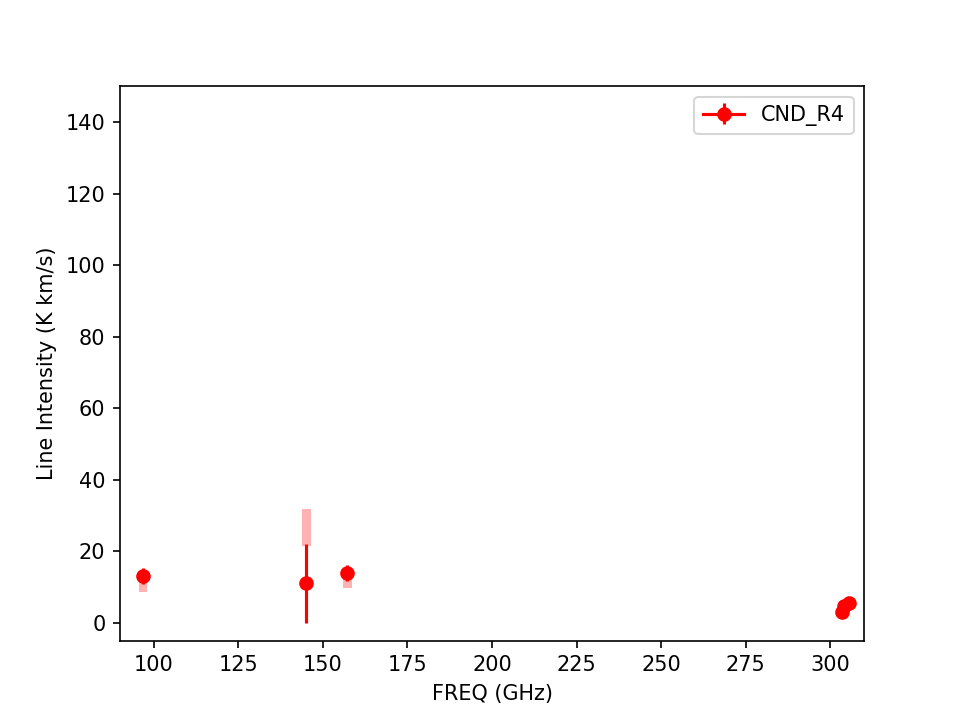} \\
    \centering\small (d) Observed and PPC intensities of \ce{CH3OH} in CND-R4
  \end{tabular}
  \caption{The {posterior predictive checks (PPCs)} of all \ce{CH3OH} intensities for four CND regions (R1-R4). {The solid circles with the thin vertical lines are the observed line intensities and the respective error bars. The shaded, thicker vertical bars are the predicted line intensities. }}
  \label{fig:PPC_CH3OH}
\end{figure*}

The comparisons are shown in Figure \ref{fig:PPC_CH3OH}. 
\renewcommand{\arraystretch}{1.0}
It is worth noting that in the 145 GHz group, the predicted line intensities are systematically higher than the observed {intensities} after correcting the line contamination from \ce{c-C3H2} emission, based on a very conservative estimate (see discussion in Sect. \ref{sec:mom0}. )
This could suggest that this correction may be too aggressive so that it underestimates the true emission contribution from \ce{CH3OH}. 
Yet owing to the gas property inference being based on multi-line \ce{CH3OH} observations, the gas properties inferred are expected to be not penalized by this fact. This characteristic also shadows the true line intensity of \ce{CH3OH} in the 145 GHz group through the pitched line intensity prediction of this group. 
We have verified the above with a separate set of gas property inference with no contamination regarding \ce{c-C3H2}, and the inferred gas properties remain consistent with the case presented in this work (i.e. with aggressive correction), and the observed versus predicted line intensity are consistent. 

\end{appendix}
%
%
\end{document}